\documentclass[twocolumn,showpacs,preprintnumbers,amsmath,amssymb,prb]{revtex4-1}
\newcommand{\etal}{{\it et al.}}

\usepackage{graphicx}
\usepackage{epstopdf}
\usepackage{dcolumn}
\usepackage{rotating}
\usepackage{setspace} 
\usepackage{amsbsy} 
\usepackage{hyperref}
\begin{document}

\title{Wavefunction Vortex Attachment via Matrix Products: Application to Atomic Fermi Gases in Flat Spin-Orbit Bands}

\author{V.W. Scarola}
\affiliation{Department of Physics, Virginia Tech, Blacksburg, Virginia 24061 USA}

\date{\today}

\begin{abstract}
Variational wavefunctions that introduce zeros (vortices) to screen repulsive interactions are typically difficult to verify in unbiased microscopic calculations. An approach is constructed to insert vortices into ansatz wavefunctions using a matrix product representation. This approach opens the door to validation of a broad class of Jastrow-based wavefunctions. The formalism is applied to a model motivated by experiments on ultracold atomic gases in the presence of synthetic spin-orbit coupling. Validated wavefunctions show that vortices in atomic Fermi gases with flat Rashba spin-orbit bands cluster near the system center and should therefore be directly visible in time-of-flight imaging.  
\end{abstract}


\maketitle

\section{Introduction}

Jastrow factors in variational wavefunctions enable the tuning of the location of wavefunction vortices to capture the essential properties of many-body problems \cite{jastrow:1955}.  While this approach has been highly successful in quantum chemistry and related fields \cite{anderson:1976,reimann:2002,saarikosk:2010}, it has also seen some success in solving flat band Hamiltonians of the form:  
 \begin{equation}
H_{\text{FB}}=\mathcal{P}V_{\text{int}}\mathcal{P},
\label{generalflatbandH}
\end{equation}
where $\mathcal{P}$ projects the inter-particle interaction, $V_{\text{int}}$, into a flat single-particle band.  The best known problems in this class include models of the quantum Hall (QH) effects \cite{laughlin:1983,jain:1989,moore:1991,booksqhe}.  But Eq.~(\ref{generalflatbandH}) describes a broad array of other compelling problems as well, e.g., electrons in graphene-based nanostructures \cite{nakada:1996,wang:2011}, atomic gases under fast rotation \cite{cooper:2008}, atoms in kagome optical lattices \cite{kagome}, fractional Chern insulators \cite{neupert:2011}, bosons in certain frustrated lattices \cite{sedrakyan:2013}, and spin-orbit coupled (SOC) systems \cite{zhou:2013,sedrakyan:2012,zhang:2013,lin:2014}.  If $\mathcal{P}$ projects onto a basis with non-commuting density operators, interesting quantum liquids or other states may arise.  But solving such problems can be prohibitive.  The non-perturbative nature of Eq.~(\ref{generalflatbandH}) implies reliance on numerical validation of ansatz wavefunctions to rigorously define quantitatively accurate solutions.  

A well known validation procedure, used in the QH regime \cite{booksqhe} to study models in the form of Eq.~(\ref{generalflatbandH}), employs parent Hamiltonians \cite{haldane:1983}.  Under this procedure a proposed wavefunction must be shown to be generated from a parent Hamiltonian and then compared with states of a physically motivated model using wavefunction overlaps and energetics.  For example, the vortex attachment protocol defined by the Laughlin state \cite{laughlin:1983} was validated with a parent Hamiltonian \cite{haldane:1983,booksqhe}.  But this process is prohibitive when studying wavefunctions that do not have simple parent Hamiltonians.  A generic wavefunction validation procedure would overcome this key difficulty.   
 
Matrix product states \cite{mps} (or, more generally, tensor networks) offer numerically efficient representations of wavefunctions that might serve as alternatives to parent Hamiltonians in validation.  Matrix product representations have been used to study entanglement in some QH states \cite{iblisdir:2007}.  A general recipe for vortex attachment in terms of matrix products would also offer a useful tool to validate ansatz states since matrix product states are straightforward to work with and related algorithms (e.g., the density matrix renormalization group \cite{white:1992,schollwoeck:2005} and other methods \cite{cirac:2009}) offer considerable opportunity for scale-up.  Promising work along these lines demonstrates a projective construction of Jastrow factors to compute local averages with local tensor networks \cite{beri:2011}.

I construct \emph{a matrix-product representation of Jastrow-based wavefunctions that offers direct validation of vortex attachment protocols} and avoids using parent Hamiltonians.  The wavefunctions constructed here use the composite fermion ansatz, originally constructed for use in the fractional QH regime \cite{jain:1989},  but can be directly compared to numerics on other models, not just QH models.  Based on a formal connection with the QH regime, I expect that vortex attachment should be useful in solving an example flat band problem motivated by recent experiments on ultracold atomic gases \cite{spinorbit}.  These experiments use lasers to generate synthetic SOC.  Assuming that synthetic Rashba SOC can be taken to the flat band limit, I find that interactions between fermions favor the clustering of vortices near the system center (similar to what has been found in studies of rotating quantum gases, quantum dots, and bosons with SOC \cite{saarikosk:2010,wang:2010,sinha:2011,ho:2011,hu:2012,zhou:2013}) and should be experimentally observable if flat band limits can be reached. 

The paper is organized as follows.  Section~\ref{sec_vortex_attach} discusses a general class of first quantized wavefunctions based on the composite fermion ansatz \cite{jain:1989}.  These wavefunctions use Jastrow factors to insert vortices to minimize interaction energy.  
Section~\ref{sec_translation_filling} describes key aspects of these wavefunctions: basis state translation and filling.  The central result of the paper is then presented in Section~\ref{sec_mps_vortex_attach}.  Here the wavefunctions are written in second quantized form and are recast in the context of matrix products.  Section~\ref{sec_apply_laughlin} uses the matrix product form for the Jastrow factor to rewrite a relevant example, the Laughlin wavefunction \cite{laughlin:1983}.  Sections~\ref{sec_appl_soc} and ~\ref{sec_numerical_results} study the flat band limit of two-dimensional fermions in the presence of parabolic trapping, Rashba SOC, and slow rotation.  Here a demonstration of the wavefunction validation process shows that interactions favor placing vortices at the center of the system instead of placing vortices on each particle, as in Laughlin-type states.  Section~\ref{sec_interprestation_observables} discusses a physical interpretation of these results and possible observables in ultracold atomic gas experiments with fermions.  

\section{Vortex Attachment in First Quantization} 
\label{sec_vortex_attach}

 I consider ground and excited state wavefunctions based on the composite fermion ansatz \cite{jain:1989}:
\begin{eqnarray}
\psi_{\nu}({\bf r}_{1},...,{\bf r}_{N}) &=& \langle {\bf r} \vert \mathcal{J_{\gamma}}^{\hspace{-0.1cm}2p}\vert \Phi_{\nu^{*}} \rangle 
\label{firstqauntizedground}
\end{eqnarray}
where the Jastrow factor $\mathcal{J}^{2p}_{\gamma}$ places $2p$ vortices in the constituent wavefunctions, $\Phi_{\nu^{*}}$, of $N$ particles.  The number of particles per basis state (the filling) in the constituent state is $\nu^{*}$.  It is convenient to take $\Phi_{\nu^{*}}$ to be a weakly interacting state, e.g., a single Slater determinant.  

Vortex insertion changes the filling because it is equivalent to removing a particle from a basis state.  $\mathcal{J}^{2p}_{\gamma}$ adds $2p$ vortices and therefore changes the number of basis states from $N\nu^{*-1}$ to $N(\nu^{*-1}+2p)$.  Section~\ref{sec_translation_filling} shows that $\psi_{\nu}$ is a state with $\nu=\nu^{*}/(2p\nu^{*}+1)$ particles per basis state.   

The following first-quantized Jastrow factor inserts $2p$ vortices independent of basis \cite{wang:2011}:
\begin{eqnarray}
\mathcal{J_{\gamma}}^{\hspace{-0.1cm}2p}=\prod_{j<k,\{\Lambda \}}^{N}\left ( T^{\dagger}_{j} - \gamma T^{\dagger}_{k}\right )^{2p}, 
\label{firstquantizedjastrow}
\end{eqnarray}
where $T^{\dagger}$ translates single-particle basis states through Hilbert space by increasing basis state indices within a set $\Lambda$.  For a basis sorted with a single index, $n$, $\Lambda$ corresponds to a 1D graph.  The Hilbert space translation operator then becomes a ladder operator\cite{girvin:1983} in $\mathcal{J}$: $T^{\dagger} \phi_{n} \propto \phi_{n+1}$, where $\phi_{n}$ is a single-particle basis state.  Section~\ref{sec_translation_filling} discusses examples of basis state translation.

The variational parameter $\gamma$ controls two types of vortex insertion.  $\gamma=1$ attaches $2p$ vortices to each particle thus lowering repulsive interaction energy by separating particles pair-wise.  States with $\gamma=0$ insert $2pN$ vortices on the $n=0$ basis state to lower occupancy of this basis state.  Wavefunctions with either $\gamma=0$ or $\gamma=1$, will be studied below.

Trial wavefunctions using Eq.~(\ref{firstquantizedjastrow}) should be energetically favorable in problems with QH features:  1) $\phi_{n}$ constitutes a flat band, 2) $\phi_{n}$ are only quasi-localized (the density operators between different $n$ do not commute), and 3) The interaction energy between basis states decreases while separating them in the graph, $\Lambda$.  These conditions describe the lowest Landau level (LLL) limit \cite{booksqhe} but can be satisfied, more generally, by other problems as well.  I will discuss strong spin orbit coupling as an additional example.

In the following I consider wavefunctions with Slater determinant constituent states $\Phi_{\nu^{*}}$.  A single Slater determinant attaches one vortex (due to Pauli exclusion) to each particle.  The following two sections will show that with this choice, Eq.~(\ref{firstqauntizedground}) reduces to the Laughlin wavefunction \cite{laughlin:1983} at $\nu=1/(2p+1)$ for $\gamma=1$ if the basis states are chosen to be LLL wavefunctions.   

\section{Basis State Translation and Filling}  
\label{sec_translation_filling}

This section shows that the operators $(T^{\dagger})^{l}$ map to translation of single-particle basis states along a graph representation of the Hilbert space.  The graph representation is used to explicitly construct a two-particle wavefunction.  The filling of $N$-particle wavefunctions constructed from $T^{\dagger}$ will also be derived explicitly.  

\begin{figure}[t]
\includegraphics[clip,width=65mm]{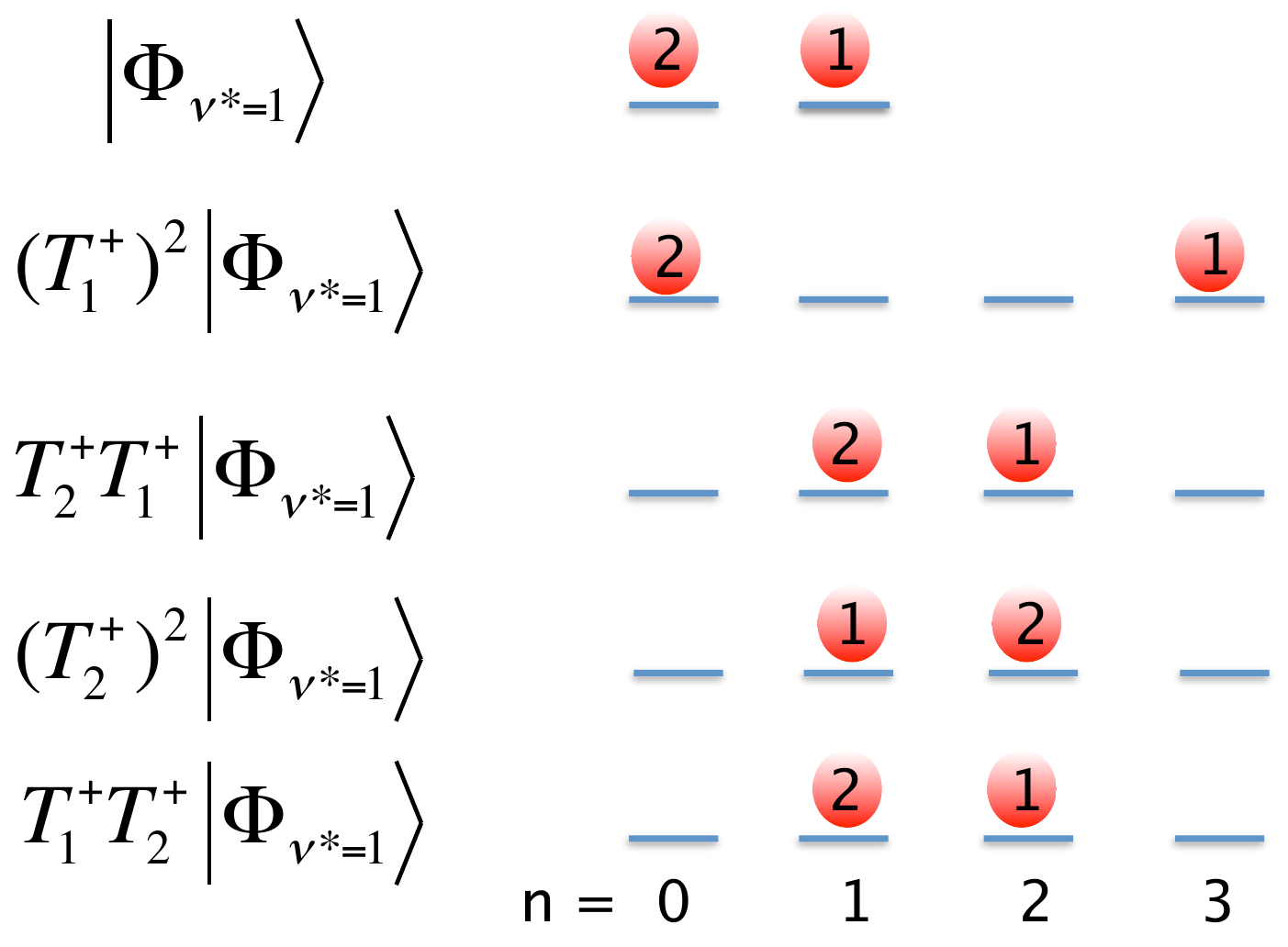}
\vspace{0in}
\caption{Schematic showing particles (circles) occupying basis states (dashes) for a two particle example of a Jastrow factor, Eq.~(\ref{supp_2jastrow}), acting on a  constituent state, Eq.~(\ref{supp_2const}).  The top row shows one of two terms in the two particle Slater determinant constituent state, Eq.~(\ref{supp_2const}).  The bottom four rows show the four configurations generated by acting Eq.~(\ref{supp_2jastrow}) on the state in the top row.  Acting Eq.~(\ref{supp_2jastrow}) on Eq.~(\ref{supp_2const}) yields four other configurations (not shown) that are the same but with the particle indices exchanged.  }
\label{supp_translation}
\end{figure}

I begin by constructing a two particle wavefunction using Eq.~(\ref{firstqauntizedground}).  I consider a degenerate Hilbert space that can be indexed with an integer: $\phi_{n}$, where $n=0,1,2,....$ labels each basis state.  Fig.~\ref{supp_translation} depicts these basis states as horizontal dashes.   

The translation operators in this case act as ladder operators on a 1D graph.   They are defined as: $T^{\dagger} \phi_{n} = f_{n} \phi_{n+1}$, where $f_{n}$ is a variational functional\cite{wang:2011} of $n$.  For simplicity I set $f_{n}=1$ in this section without loss of generality.  The operator representation allows a polynomial construction of Jastrow factors even though the basis states themselves are not polynomials, i.e., $\langle {\bf r}_{1}  \vert (T^{\dagger}_{1})^{2} \vert  \phi_{0} \rangle=  \phi_{2}({\bf r}_{1})$ but 
$ \phi_{2}({\bf r}_{1}) \neq  \phi_{1}({\bf r}_{1}) \phi_{1}({\bf r}_{1})$. $N=2$ in Eq.~(\ref{firstquantizedjastrow}) gives:
\begin{eqnarray}
\mathcal{J}_{\gamma=1}^{2}=\left ( T^{\dagger}_{1} -  T^{\dagger}_{2}\right )^{2},
\label{supp_2jastrow}
\end{eqnarray}
for $p=1$.  I use this form to construct a two-particle ansatz wavefunction below.

The operator form of the Jastrow factor allows a direct basis-independent construction of the total wavefunction.  I select a two particle determinant state as the constituent state:
\begin{eqnarray}
\langle {\bf r}_{1},{\bf r}_{2} \vert \Phi_{\nu^{*}=1}\rangle \propto \phi_{0}({\bf r}_{1})\phi_{1}({\bf r}_{2})-\phi_{1}({\bf r}_{1})\phi_{0}({\bf r}_{2}),
\label{supp_2const}
\end{eqnarray}
where the constituent filling is $\nu^{*}=1$ because there is one particle per basis state.  This state is depicted schematically in the top row of Fig.~\ref{supp_translation}.

I now construct the total $N=2$ wavefunction using Eq.~(\ref{firstqauntizedground}).  Acting Eq.~(\ref{supp_2jastrow}) on Eq.~(\ref{supp_2const}) yields:
\begin{eqnarray}
\langle {\bf r}_{1},{\bf r}_{2} \vert  \psi_{\nu=1/3} \rangle &\propto&  \phi_{3}({\bf r}_{1})\phi_{0}({\bf r}_{2}) - 3\phi_{2}({\bf r}_{1})\phi_{1}({\bf r}_{2}) 
\nonumber \\
&-&  \phi_{3}({\bf r}_{2})\phi_{0}({\bf r}_{1})+ 3\phi_{2}({\bf r}_{2})\phi_{1}({\bf r}_{1}).
\label{supp_2totalwf}
\end{eqnarray}
Here the total wavefunction results from a sum of eight terms.  Four of the terms are depicted schematically in Fig.~\ref{supp_translation}.  The other four terms result from swapping particle coordinates.  

The wavefunction written in Eq.~(\ref{supp_2totalwf}) appears to have a filling different from 1/3.  The enumeration of basis states in Fig.~\ref{supp_translation} shows that there are only four basis states for two particles.  The filling for the two-particle equation, Eq.~(\ref{supp_2totalwf}), is therefore 1/2.  But it can be shown that the filling of Eq.~(\ref{firstqauntizedground}) converges to 1/3 in the large $N$ limit for $p=1$ and $\nu^{*}=1$.

I now show that the filling of Eq.~(\ref{firstqauntizedground}) converges to $\nu^{*}/(2p\nu^{*}+1)$ in the large $N$ limit \cite{jain:1989,booksqhe}.  The number of basis states in any constituent state at filling $\nu^{*}$ is, by definition, $N\nu^{*-1}+c_{1}$, where $c_{1}$ is a constant of order unity.  If the Jastrow factor inserts $2p$ vortices per particle, each vortex vacates a basis state.  The number of basis states then increases to $N(\nu^{*-1}+2p)+c_{2}+c_{1}$, where $c_{2}$ is also a constant of order unity.  This shows that the number of particles per basis state yields a filling: $\nu=\nu^{*}/(2p\nu^{*}+1)$, in the large $N$ limit.  

The filling can also be derived explicitly from specific forms for the Jastrow factor and the constituent state.  Consider the largest power in the Jastrow factor defined in Eq.~(\ref{firstquantizedjastrow}):
$
\mathcal{J}_{\gamma}^{2p}=(T^{\dagger}_{1})^{2p(N-1)}+...
$
This shows that $\mathcal{J}_{\gamma}^{2p}$ shifts the largest basis state index by $2p(N-1)$, i.e., $c_{2}=-2p$.  A single Slater determinant constituent state can also be written in terms of translation operators:
$
\vert \Phi_{\nu^{*}=1 } \rangle \propto \prod_{j<k}^{N}\left ( T^{\dagger}_{j} - \gamma T^{\dagger}_{k}\right )\vert 0 \rangle =(T^{\dagger}_{1})^{N-1}\vert 0 \rangle+...,
$
where the last equality shows that this constituent state has a basis state index that is at most $N-1$.  Putting these two results together, the largest power of the  $T^{\dagger}_{1}$ operator in $\mathcal{J}_{\gamma}^{2p}\vert \Phi_{\nu^{*}=1 } \rangle$ is $(2p+1)(N-1)$.  The number of basis states is therefore $(2p+1)(N-1)+1$, with $c_{2}=-2p$ and $c_{1}=0$. This leaves the filling, defined as the number of particles per basis state, to be:
$
 \nu=\frac{N}{(2p+1)(N-1)+1} \underset{N\rightarrow\infty}{\rightarrow} \frac{1}{2p+1}.
$
The first equality shows that, for $N=2$ and $p=1$, the filling is $1/2$ (as shown schematically in Fig.~\ref{supp_translation}).  But the large $N$ limit yields $\nu\rightarrow 1/3$ for $p=1$.  

\section{Matrix Product Formulation for Vortex Attachment}  
\label{sec_mps_vortex_attach}

Recasting the Jastrow factors considered above in second quantization allows a representation in terms of matrix products.  In the following, first-quantized $N$-body operators $\mathcal{O}$ will be represented in second quantization by 
$\hat{\mathcal{O}}=\int \hat{\Psi}^{\dagger}({\bf r}_{1})\cdots\hat{\Psi}^{\dagger}({\bf r}_{N})\mathcal{O}\hat{\Psi}^{\vphantom{\dagger}}({\bf r}_{1})\cdots\hat{\Psi}^{\vphantom{\dagger}}({\bf r}_{N})d{\bf r}_{1}\cdots d{\bf r}_{N}$, 
where $\hat{\Psi}^{\dagger}({\bf r})$ is a fermion field operator.  Expanding the field operators imposes a specific basis choice in Fock-space: $\hat{\Psi}^{\dagger}({\bf r})=\sum_{n}\phi_{n}({\bf r})\hat{c}^{\dagger}_{n}$, where $\hat{c}^{\dagger}_{n}$ creates a fermion in the state $\phi_{n}$.  Using this expansion, Eq.~(\ref{firstquantizedjastrow}) can be rewritten as sums over $T^{\dagger}$ using Shiota's formula \cite{difrancesco:1994} that, in turn, allows a second-quantized representation (See Appendix):
\begin{eqnarray}
\hat{\mathcal{J}}_{\gamma}^{\hspace{-0.005cm}2p}&=&(-1)^{\frac{pN(N-1)}{2} }\sum_{k=1}^{N(N-1)}\frac{(-1)^{k}}{k!}\times \nonumber \\
& &\hspace{-1.5cm} \sum_{\substack{n_{1}\geq1,...,n_{k}\geq1 \\n_{1}+...n_{k}=N(N-1)}} \prod_{i=1}^{k} \frac{1}{2n_{i}}\sum_{l=0}^{n_{i}p} {n_{i}p \choose l}  (-1)^{l} \hat{M}_{n_{i}p-l}\hat{M}_{l},
\label{secondquantizedJastrow}
\end{eqnarray}
where,
\begin{eqnarray}
\hat{M}_{l}=\sum_{n,n'}\gamma_{n,n'}^{l} \hat{c}^{\dagger}_{n}\hat{c}^{\vphantom{\dagger}}_{n'}
\label{secondquantizedmatrix}
\end{eqnarray}
defines a matrix in terms of variational parameters $\gamma_{n,n'}^{l}$.  For $\gamma=1$ they are given by $\gamma_{n,n'}^{l}=\langle \phi_{n}\vert  \left( T^{\dagger}\right)^{l} \vert \phi_{n'} \rangle$.  Section~\ref{sec_appl_soc} shows that the $\gamma=0$ limit arises when the variational parameters $\gamma_{n,n'}^{l}$ are chosen such that  $\gamma_{n,n'}^{l}\propto \delta_{l,1}\langle \phi_{n}\vert  \left( T^{\dagger}\right)^{l} \vert \phi_{n'} \rangle$.

Eqs.~(\ref{secondquantizedJastrow}) and (\ref{secondquantizedmatrix}) define the centerpiece of this work.  Wavefunctions constructed from these Jastrow factors can be compared directly with model diagonalization without reference to parent Hamiltonians.  The matrix sizes in the above equations can, in some cases, exhibit a basis-dependent exponential scaling with $N$ which at first appears prohibitive.  Nonetheless, there are two situations where these relations can be used: 1)  In their exact form, Eqs.~(\ref{secondquantizedJastrow}) and (\ref{secondquantizedmatrix}) are suited to small system size validation studies.  I demonstrate the validation process below.  Once validated, the ansatz wavefunctions in their first-quantized form can be used in larger systems.  2)  Writing the Jastrow factor in terms of a product over matrices also allows use of well known approximations for optimization, implementation, and validation.  Approximate wavefunctions derived from Eq.~(\ref{secondquantizedJastrow}) using singular value decomposition will be explored in future work.

I rewrite the ansatz states, Eqs.~(\ref{firstqauntizedground}), in second quantization using Eq.~(\ref{secondquantizedJastrow}):
\begin{eqnarray}
\vert \hat{\psi}_{1/(2p+1)} \rangle&=&\hat{\mathcal{J}}^{2p}_{\gamma}\prod_{n=0}^{N-1} \hat{c}^{\dagger}_{n}\vert \hat{0}\rangle.
\label{secondquantizedlaughlin}
\end{eqnarray}
This form shows that $\mathcal{J}^{2p}_{\gamma}$ can be written as a sum of matrix products acting on a column vector defined by $\hat{\Phi}_{\nu^{*}=1}=\prod_{n=0}^{N-1} \hat{c}^{\dagger}_{n}\vert \hat{0}\rangle$.  Excited states can also be constructed using a different constituent state: $\hat{c}^{\dagger}_{N-1+\Delta M}\prod_{n=0}^{N-2}\hat{c}^{\dagger}_{n}\vert \hat{0}\rangle$, where the $n=N-1$ state increments by $\Delta M$.  I implement examples below.

\section{Application to Single Component Laughlin States}
\label{sec_apply_laughlin}

 As a useful first example I rewrite the Laughlin ground state as a product of matrices \cite{iblisdir:2007}.   A specific Jastrow factor and a specific constituent state (a Slater determinant) are chosen and inserted into Eq.~(\ref{firstqauntizedground}).  This yields a specific operator form for the trial state.  By choosing LLL single-particle basis states in the symmetric gauge, the familiar form for the first quantized Laughlin wavefunction is recovered.  I then compute the matrix elements used to rewrite the Laughlin wavefunction with Eq.~(\ref{secondquantizedJastrow})
 
I start with a specific operator form for the Jastrow factor: 
\begin{eqnarray}
\mathcal{J}_{\gamma=1}^{\hspace{0.0cm}2p}=\prod_{j<k}^{N}\left ( T^{\dagger}_{j} - T^{\dagger}_{k}\right )^{2p}, 
\label{supp_firstquantizedjastrow}
\end{eqnarray}
and a specific constituent state:
\begin{eqnarray}
\vert \Phi_{\nu^{*}=1} \rangle\propto \prod_{j<k}^{N}\left ( T^{\dagger}_{j} -  T^{\dagger}_{k} \right ) \vert 0 \rangle,
\label{supp_firstquantizedconstituent}
\end{eqnarray}
where the vacuum is defined by $\langle {\bf r}  \vert 0 \rangle=\prod_{j} \phi_{n=0}({\bf r}_{j})$.  The constituent state $\vert \Phi_{\nu^{*}=1} \rangle$ is equivalent to a Slater determinant.  Substituting Eqs.~(\ref{supp_firstquantizedjastrow}) and ~(\ref{supp_firstquantizedconstituent}) into Eq.~(2) leads to an operator form for the Laughlin state \cite{girvin:1983,wang:2011}:
\begin{eqnarray}
\vert \psi_{\nu=1/(2p+1)} \rangle&=&\prod_{j<k}^{N}\left ( T^{\dagger}_{j} - T^{\dagger}_{k}\right )^{2p}\vert \Phi_{\nu^{*}=1} \rangle \nonumber \\
&\propto& \prod_{j<k}^{N}\left ( T^{\dagger}_{j} - T^{\dagger}_{k}\right )^{2p+1} \vert 0 \rangle.
\end{eqnarray}
Here the wavefunctions are written in terms of basis-independent Hilbert space translation operators.  To connect with more familiar forms for the Laughlin state, a basis must be chosen.   

\begin{figure}[t]
\includegraphics[clip,width=85mm]{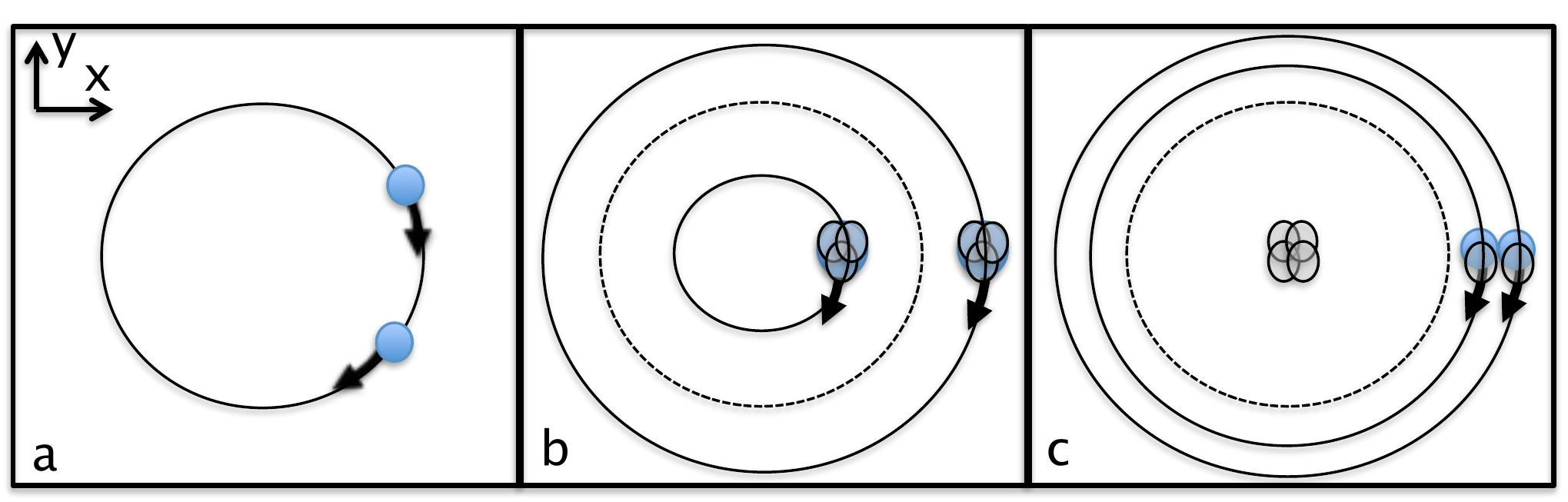}
\vspace{0in}
\caption{(Color online) a) Schematic of two particles in an angular momentum eigenstate in the disk geometry. Rings set an average inter-particle spacing.  b) Attaching one wavefunction vortex to each particle accounts for Pauli exclusion.  Attaching two additional vortices ($\gamma=1$ and $p=1$ in Eq.~(\ref{firstquantizedjastrow})) separates them further.  c) Placing the two additional vortices at the center ($\gamma=0$ and $p=1$ in Eq.~(\ref{firstquantizedjastrow})) forces occupancy of higher angular momenta.  }
\label{vortices}
\end{figure}

The Laughlin wavefunction was originally constructed in the symmetric gauge appropriate for a disk geometry.  The LLL basis states in this gauge are \cite{laughlin:1983,booksqhe}:
$\phi^{\text{D}}_{m}=z^{m}(2\pi 2^{m}m!)^{-1/2}\exp(-\vert z \vert^{2}/4)$, in units of the magnetic length, where $z=x-iy$ is a complex planar coordinate and $m=0,1,...$ indexes angular momentum.  The translation operators in this basis give: $\langle {\bf r}  \vert (T^{\dagger})^{l} \vert  \phi_{0} \rangle=f_{l} \sqrt{2^{l}l!}\phi^{\text{D}}_{l}({\bf r})=(2\pi)^{-1/2}z^{l}\exp(-\vert z \vert^{2}/4)$.  The choice $f_{l}=1$ yields the Laughlin wavefunction:
\begin{eqnarray}
\langle {\bf r} \vert \psi_{\nu=\frac{1}{2p+1}} \rangle \propto \prod_{j<k}^{N}\left ( z_{j} - z_{k}\right )^{2p+1} \prod_{i=1}^{N} e^{-\vert z_{i} \vert^{2}/4}.
\end{eqnarray}
This shows that the familiar form for the Laughlin state follows from Eq.~(\ref{firstqauntizedground}) with specific choices for the Jastrow factor, the constituent state, and basis.   Fig.~\ref{vortices} shows a schematic for vortex attachment in this basis.

The Laughlin state can now be written in term of matrix products using Eq.~(\ref{secondquantizedJastrow}).  To compute the matrix elements $\gamma_{n,n'}^{l}$ I require the translation operator to act as a polynomial in $z$.  This can be achieved using   
$m\rightarrow n$ to map the problem to single-particle basis states on a 1D graph labeled by $n$.  Using $\int d^{2}r \phi^{\text{D}}_{n}z^{l}\phi^{\text{D}}_{n'}=\Gamma_{n'}^{l}\delta_{n,n'+l}$, leads to:
\begin{equation}
\left. \begin{array}{lcl}
\gamma_{n,n'}^{l}&\rightarrow& \Gamma_{n'}^{l}\delta_{n,n'+l} 
\end{array}\right\}  
\begin{array}{l} \text{Laughlin in }\\
 \text{Disk Basis}.
 \end{array}
\label{laughlinmatrices}
\end{equation}
 where $\Gamma_{n}^{l}$ is:
\begin{eqnarray}
\Gamma_{n'}^{l}\equiv\sqrt{\frac{2^{l}(n'+l)!}{n'!}}.
\label{laughlingamma}
\end{eqnarray}

This choice for the matrix elements completely specifies the Laughlin state in second quantization [using Eq.~(\ref{secondquantizedJastrow}), Eq.~(\ref{secondquantizedlaughlin}), and (\ref{laughlinmatrices})] and defines a matrix product representation equivalent to those in Ref.~\onlinecite{iblisdir:2007}.  I have checked that the wavefunction amplitudes specified above reproduce the Laughlin state obtained from the parent Hamiltonian \cite{haldane:1983}.

\section{Application to Spin-Orbit Coupled Fermions} 
\label{sec_appl_soc}

In this section I now turn to a flat band problem with non-polynomial basis states motivated by ultracold atomic gas experiments \cite{spinorbit}.  In such experiments external lasers can be used to confine fermionic alkali atoms, e.g., $^{40}$K, to 2D.  Two hyperfine levels define a pseudo-spin.  The laser beam waist induces a parabolic trapping while applied Raman beams have recently demonstrated the ability to apply synthetic spin-obit coupling \cite{spinorbit}.  I first discuss single-particle properties of a model of Rashba SOC.  I then model interactions within a flat single-particle band.  

\begin{figure}[t]
\vspace{-0.2in}
\includegraphics[clip,width=90mm]{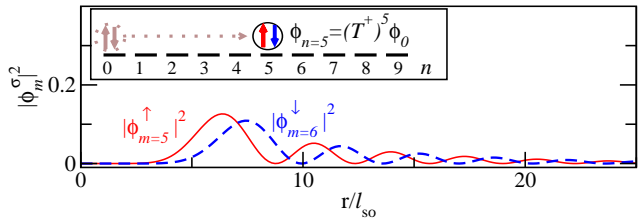}
\vspace{-1.5in}
\caption{(Color online) The single-particle basis states as a function of position for each spin component in Eq.~(\ref{phibasis}) for $\alpha=100$.  The inset depicts the mapping between angular momentum, indexed by $m$, and a 1D graph indexed by $n$.}
\label{basis}
\end{figure}

I begin with the full 2D model of two-component fermions \cite{sinha:2011,hu:2012,sedrakyan:2012,sedrakyan:2013,zhou:2013}:
\begin{eqnarray}
&\hat{H}&=\int d^{2}{\bf r} \hat{\Psi}^{\dagger}_{\sigma}({\bf r})\left [ 
-\frac{\hbar^{2}\nabla^{2}}{2m_{p}}+\frac{m_{p}\omega_{\text{T}}r^{2}}{2} \right.\nonumber\\
&+&\left.  
\hbar \lambda i \left ( \partial_{x}\sigma_{y}- \partial_{y}\sigma_{x}
 \right ) 
 +\frac{g_{2D}^{\sigma,\sigma'}}{2} \hat{\Psi}^{\dagger}_{\sigma'}({\bf r})\hat{\Psi}^{\vphantom{\dagger}}_{\sigma'}({\bf r})\right ]\hat{\Psi}^{\vphantom{\dagger}}_{\sigma}({\bf r}) \hspace{0.2cm}
 \label{spinorbitH}
\end{eqnarray}
where the field operator $\hat{\Psi}^{\dagger}_{\sigma}$ creates a fermion at ${\bf r}= (x,y)$ in spin state $\sigma=\uparrow,\downarrow$, $\omega_{\text{T}}$ specifies the trapping frequency due to a parabolic confinement of particles of mass $m_{p}$, $\lambda$ is the strength of the Rashba term, and $\boldsymbol{\sigma}$ are the Pauli matrices.  For fermions, the interaction term originates from a contact interaction, $g_{2D}^{\uparrow,\downarrow}\delta({\bf r}-{\bf r}')$, with strength $g_{2D}^{\uparrow,\downarrow}=\sqrt{8\pi}\hbar^{2}a_{s}/m_{p}l_{z}$, where $l_{z}$ is the harmonic confinement length along the $z$-direction and $a_{s}$ is the $s$-wave scattering length.  

I first discuss the non-interacting limit ($g_{2D}^{\sigma,\sigma'}=0$).  The single-particle part of the Hamiltonian can be rewritten: 
$
H_{0}=(-i\hbar\nabla+\lambda m_{p}\hat{z}\times{\boldsymbol \sigma})^{2}/2m_{p}+m_{p}\omega_{\text{T}}r^{2}/2.
$
This form shows that the Rashba term appears as a non-Abelian vector potential that suggests a magnetic field-like interpretation. 

Recent work \cite{hu:2012,sedrakyan:2012,zhou:2013} shows that low energy single-particle basis states of trapped Rashba particles define a flat band that resembles the LLL for strong SOC.  For $\omega_{T}=0$ the eigenstates of $H_{0}$ define a ``Mexican-hat'' potential in momentum space.  But the trapping gaps out all but one low energy ring that forms a flat band for $\alpha \gg 1$, with $\alpha\equiv l_{\text{T}}/l_{\text{SO}}$, $l_{\text{T}}\equiv(\hbar/m_{p}\omega_{T})^{1/2}$, and $l_{\text{SO}}\equiv\hbar/m_{p}\lambda$.  The  degenerate single-particle energies, $\approx (1-\alpha^{2}/2)\hbar \omega_{\text{T}}+\mathcal{O}(m^{2}/\alpha^{2})$, contribute just to the zero-point energy.  This shows that the flat band limit is a good approximation for $m<\alpha$.

The resulting single-particle basis states defining the lowest energies are well approximated\cite{zhou:2013} by spinor eigenstates of total angular momentum, $L_{z}+\hbar\sigma_{z}/2$, with eigenvalues $\hbar(m+1/2)$:
\begin{eqnarray}
\phi_{m}(r,\theta)=\frac{\exp (-r^{2}/2\alpha^{2})}{N_{m}}\begin{pmatrix}
  e^{ i m \theta} J_{m}(r)\\
   e^{ i (m+1) \theta} J_{m+1}(r)
\end{pmatrix},
\label{phibasis}
\end{eqnarray}
where $N_{m}^{2}\equiv\pi\alpha^{2}\exp(-\alpha^{2}/2)[I_{m}(\alpha^{2}/2)+I_{m+1}(\alpha^{2}/2) ]$ defines the normalization, $J_{m}$ ($I_{m}$) are the Bessel (modified Bessel) functions, and $l_{\text{SO}}$ is the unit of length. The $\phi_{m}(r,\theta)$ define a \emph{helicity basis} because these states are also eigenstates of the helicity operator, ${\boldsymbol \sigma}\cdot {\boldsymbol L}$, where $\boldsymbol{L}$ is the angular momentum operator.

Adding the following term: $-\omega_{\text{T}} \sigma_{z}L_{z}$, to $H_{0}$ yields Eq.~(\ref{phibasis}) as exact eigenstates \cite{zhou:2013}.  This simplifies calculation of interaction matrix elements but the term is not straightforward to generate in atomic gas experiments.  It can be shown \cite{zhou:2013} that, in the absence of a $-\omega_{\text{T}} \sigma_{z}L_{z}$ term, Eq.~(\ref{phibasis}) is still a good approximate solution to $H_{0}$ for $\alpha \gg 1$. 

Eq.~(\ref{spinorbitH}) is time-reversal invariant.  As a result, the single-particle basis states $\phi_{m}$ belong to a Kramers degenerate pair, the other member having the opposite angular momentum.  The ansatz wavefunctions discussed here can be generalized to a two-component basis but as a first test I restrict the basis to a single component by breaking time reversal symmetry with slow rotation.  

Eq.~(\ref{spinorbitH}) can be written in a rotating frame of reference.  Under slow rotation I  include a term: $-\omega_{R}L_{z}$, to  impose a splitting $\sim\hbar \omega_{\text{R}} m$ between the Kramers degenerate pairs.  Dynamical corrections $\sim\omega_{R}^{2}$ can be ignored for slow rotation \cite{radic:2011}.  Here I also assume that rotation induced Zeeman terms discussed in Ref.~\onlinecite{zhou:2013} are canceled with an applied Zeeman coupling.  In this limit the spinors $\phi_{m}$ form a basis in a degenerate kinetic energy band at fixed total angular momentum, $M\equiv\sum_{i=1}^{N}m_{i}$, for $ 0\leq m<\alpha$ and $\alpha \gg1$.  

Eq.~(\ref{phibasis}) defines a quasi-localized basis set that approximate LLL functions for $r\rightarrow0$.  For $\alpha\ll 1$ the basis states are Gaussians in $r$ but for  $\alpha\gg 1$ the Bessel function imposes an oscillating tail. Fig.~(\ref{basis}) shows the peak of $\phi_{m}$ increasing along $r$ as $m$ increases.  

I now consider interactions, $g_{2D}^{\uparrow,\downarrow}>0$ in Eq.~(\ref{spinorbitH}).  
Requiring $m<\alpha$ and $\alpha \gg1$ ensures the flat band limit. $s$-wave scattering dominates interactions between ultracold alkali atoms.  I therefore use $V_{\text{int}} \rightarrow g_{2D}^{\uparrow,\downarrow} \delta( {\bf r}-{\bf r}' )$, where $g_{2D}^{\uparrow,\downarrow}$ is an experimentally tunable interaction strength between fermions of opposite spin \cite{vyasanakere:2011}, in Eq.~(\ref{generalflatbandH}) with $\mathcal{P}$ projecting into $\phi_{n}$.  Projection follows from an expansion in the flat band basis:
$
\hat{\Psi}^{\dagger}=\sum_{m}\phi_{m} \hat{c}^{\dagger}_{m},
$
where $\hat{c}^{\dagger}_{m}$ creates a fermion in the helicity eigenstate defined by Eq.~(\ref{phibasis}).  
 I study the spectrum at fixed $M$.  Working with fixed $M$ corresponds to a point along the Yrast line \cite{saarikosk:2010}.  
For fixed $M$, the interaction then becomes the only non-constant term, thus leaving a Hamiltonian in the form of Eq.~(\ref{generalflatbandH}):
\begin{equation}
\hat{H}\approx\frac{g_{2D}^{\uparrow,\downarrow}}{2}
\sum_{\{m\}}\langle \phi_{m_{1}}, \phi_{m_{2}} \vert  \phi_{m_{3}}, \phi_{m_{4}}\rangle
\hat{c}^{\dagger}_{m_{1}}\hat{c}^{\dagger}_{m_{2}}\hat{c}^{\vphantom{\dagger}}_{m_{3}}\hat{c}^{\vphantom{\dagger}}_{m_{4}},
\label{spinorbithfb}
\end{equation}
where the sum is over allowed indices, $m_{1}+m_{2}=m_{3}+m_{4}$.  This shows that an interaction-only flat band model, Eq.~(\ref{generalflatbandH}), derives from $\hat{H}$.  Eq.~(\ref{spinorbithfb}) also shows the form of the interaction matrix used to numerically diagonalize $g_{2D}^{\uparrow,\downarrow}\delta({\bf r}-{\bf r}')$ in the $\phi_{m}$ basis. 

I now discuss possible wavefunctions designed to capture the essential properties of the eigenstates of  Eq.~(\ref{spinorbithfb}).  Fig.~(\ref{basis}) shows that spatially decaying interactions should decrease in strength as $|m-m'|$ increases. This suggests that here Eq.~(\ref{firstquantizedjastrow}) [or, equivalently, Eq.~(\ref{secondquantizedJastrow}) with $m\rightarrow n$ ] will offer an energetically favorable method to impose vortex attachment because: 1) the basis defines a flat band, 2) the basis states are only quasi-localized and therefore define non-commuting density operators, and 3) Hilbert space translation spatially separates particles to decrease the interaction energy.

To capture the effects of interactions I consider wavefunctions written at $\nu=1/(2p+1)$ which correspond to ground states at $M=(2p+1)N(N-1)/2$ and $2p+1$ vortices per particle.  I use the formalism introduced above [Eqs.~(\ref{secondquantizedJastrow})-(\ref{secondquantizedlaughlin})] to write ansatz states in the matrix product form.  I consider two different states (one with $\gamma=1$ and one with  $\gamma=0$)  in the helicity basis.

A $\gamma=1$ Laughlin-type state in the helicity basis can be derived from the $r\rightarrow 0$ limit of Eq.~(\ref{phibasis}).    I use the definition of the matrices,  $\gamma_{n,n'}^{l}=\langle \phi_{n}\vert  \left( T^{\dagger}\right)^{l} \vert \phi_{n'} \rangle$, in terms of single-particle basis states.  This can be done numerically for the basis states defined by $\phi_{m}$ but an analytic expression is possible by noting that $\phi_{m}$ defines a LLL-like Hilbert space of spinors.  Using this I derive analytic expressions for  $\gamma_{n,n'}^{l}$ in the helicity basis. 

I first note that delta function interactions will emphasize the short range part of the basis states.  One can show that, in the $r\rightarrow 0$ limit,  the upper and lower entries in the spinor defining $\phi_{m}$ reduce to lowest Landau level basis states, $\phi^{\text{D}}_{m}$ and $\phi^{\text{D}}_{m+1}$:
\begin{equation}
\phi_{m}\rightarrow \frac{1}{\sqrt{2}} \begin{pmatrix}
  \phi_{m,\uparrow}^{D}\\
    \phi_{m+1,\downarrow}^{D}
\end{pmatrix},
\label{spinorlimit}
\end{equation}
where I have taken the limit $r\rightarrow 0$ with $r/\alpha$ held constant.  This simplification allows use of the lowest Landau level basis states to define the matrix elements with: $\left( T^{\dagger}\right)^{l} \vert \phi_{m'}^{D} \rangle= f_{m'} \Gamma_{m'}^{l} \vert \phi_{m'+l}^{D} \rangle$, where $f_{m'}$ is a variational functional of $m'$.  

Using Eq.~(\ref{spinorlimit}) I find expressions for $\gamma_{n,n'}^{l}$ for the Laughlin state in the helicity basis:
\begin{eqnarray}
&&\langle \phi_{m}\vert  \left( T^{\dagger}\right)^{l} \vert \phi_{m'} \rangle  \nonumber \\ 
&\rightarrow& \frac{\langle  \phi_{m,\uparrow}^{D} \vert  \left( T^{\dagger}\right)^{l} \vert \phi_{m',\uparrow}^{D} \rangle +
\langle  \phi_{m+1,\downarrow}^{D} \vert  \left( T^{\dagger}\right)^{l} \vert \phi_{m'+1,\downarrow}^{D} \rangle }{2}  \nonumber \\ 
&=& \frac{(f_{m',\uparrow}\Gamma_{m'}^{l}+f_{m'+1,\downarrow}\Gamma_{m'+1}^{l})\delta_{m,m'+l}}{2}, 
\label{matrixelement}
\end{eqnarray}
where the arrow indicates the limit $r\rightarrow 0$ with $r/\alpha$ held constant. This expression allows a definition of the matrix elements that define the Laughlin state in the helicity basis:
\begin{equation}
\begin{split}
\hspace{0cm}
\left. \begin{array}{lcl}
\gamma_{n,n'}^{l}&\rightarrow&(\Gamma_{n'}^{l}+\Gamma_{n'+1}^{l})\delta_{n,n'+l}  \\ 
\end{array}\right\}
\begin{array}{l} \text{Laughlin in }\\
 \text{Helicity Basis}
 \end{array}
\end{split}
\label{laughlinjastrow}
\end{equation}
by setting $f_{m',\uparrow}=f_{m',\downarrow}=2$ and $m\rightarrow n$.  This choice for $f_{m}$ simply adjusts the normalization.  $m$ dependence in $f_{m}$ impacts energetics.

Eq.~(\ref{matrixelement}) can also be used to derive the matrix elements for the central vortex state (defined by $\gamma=0$).  In setting $\gamma=0$ I note that the only terms that survive in Eq.~(\ref{secondquantizedJastrow}) have $l=1$.  This shows that Eq.~(\ref{matrixelement}) gives:
\begin{equation}
\begin{split}
\hspace{-0.5cm}
\left. \begin{array}{lcl}
\gamma_{n,n'}^{l}&\rightarrow&(\Gamma_{n'}^{l}+\frac{\Gamma_{n'+1}^{l}}{\sqrt{n+1}})\delta_{n,n'+l}\delta_{l,1}   \\ 
\end{array}\right\} 
\begin{array}{l} \text{Central Vortex }\\
 \text{in Helicity Basis}
\end{array}
\end{split}
\label{corejastrow}
\end{equation}
with $l=1$, $m\rightarrow n$, $f_{m',\uparrow}=2$, and $f_{m'+1,\downarrow}=2/\sqrt{m'+1}$.  This choice for $f_{m',\downarrow}$ prevents the density from vanishing at the center and was found to give the best overlaps in numerics.  

When these Jastrow factors are inserted into Eq.~(\ref{secondquantizedlaughlin}) they offer distinct ansatz states that can be directly compared with the results of exact diagonalization.  Eq.~(\ref{laughlinjastrow}) places three vortices on each particle ($\gamma=1$) to build the Laughlin state in the helicity basis, Eq.~(\ref{phibasis}).  But Eq.~(\ref{corejastrow}) places $2N$ vortices in the system center and one on each particle ($\gamma=0$) in the helicity basis. A numerical routine (provided in EPAPS \cite{epaps}) uses the simplicity of the matrix product representation and matrix multiplication methods on sparse matrices to generate these states.

\begin{figure}[t]
\includegraphics[clip,width=85mm]{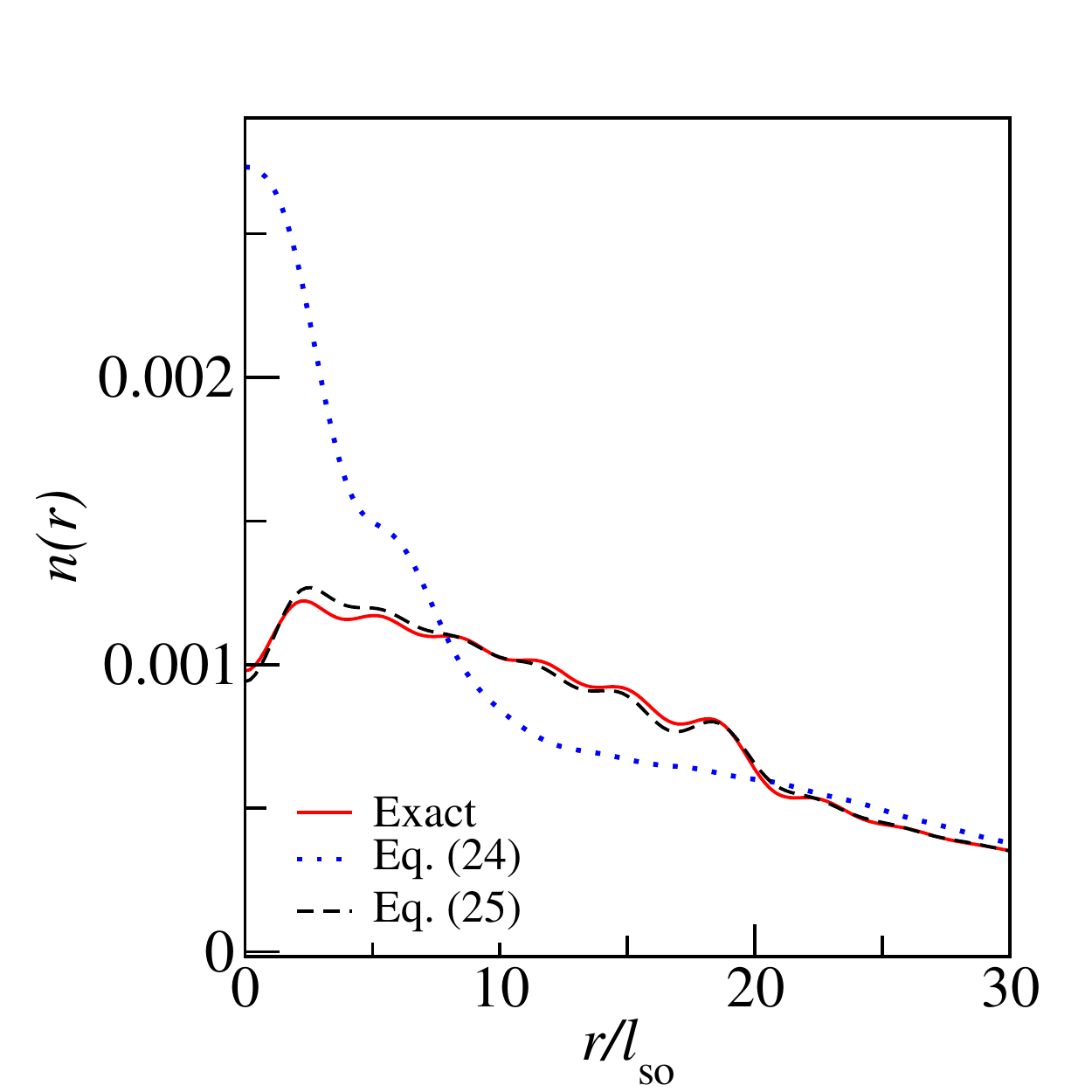}
\vspace{-2.9in}

\includegraphics[clip,width=1.76in]{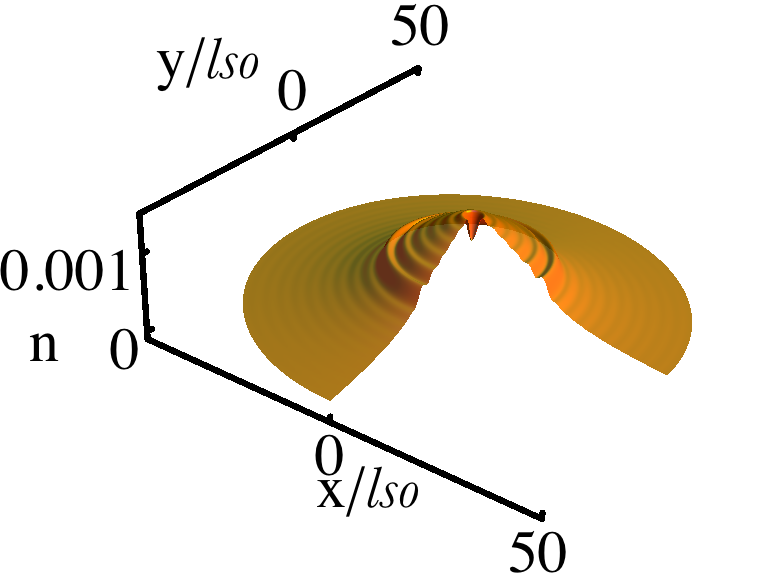}
\hspace{-1.1in}
\vspace{1.7in}

\caption{(Color online)  Main Panel: Density as a function of radial position for $\alpha=100$, $N=6$ particles, and total angular momentum $M=45$ which corresponds to $\nu=1/3$. The solid line results from diagonalization of the interaction-only model, Eq.~(\ref{spinorbithfb}).  The dotted [dashed] line was computed using Eqs.~(\ref{secondquantizedlaughlin}) and (\ref{laughlinjastrow})  [Eqs.~(\ref{secondquantizedlaughlin}) and (\ref{corejastrow})] with $p=1$. Inset: the same as the dashed line but versus $x$ and $y$.}
\label{density_comparison}
\end{figure}

\section{Numerical Results} 
\label{sec_numerical_results}

 I illustrate validation by comparing results from numerical diagonalization of Eq.~(\ref{spinorbithfb}) with ansatz wavefunctions defined by Eqs.~(\ref{laughlinjastrow}) and (\ref{corejastrow}) for $p=1$.

Fig.~\ref{density_comparison} compares the density of the ansatz states with the exact state.  The Laughlin state describes a state with uniform density near the trap center for large $N$.  The central vortex state, by contrast, shows a pronounced dip, detailed in the inset.  The density comparison shows good agreement between the exact state and central vortex state. The overlaps and energies in Table~\ref{tab1} also show good agreement.

Decreasing $\alpha$ induces a transition from the central vortex state to a more uniform state.  At smaller $\alpha$, Eq.~(\ref{laughlinjastrow}) has a higher overlap (at most $0.5$ at $\alpha\approx 23$) because in this limit the basis states better approximate $\phi^{\text{D}}$.  Here interactions appear to favor vortex attachment on each particle.  But the limit $\alpha\sim100$ is consistent with the flat band assumption, $m<\alpha$.  The central vortex state, Eq.~(\ref{corejastrow}), therefore captures the low energy physics for the physically relevant limit, $\alpha > 23$.

\section{Interpretation and Observables}
\label{sec_interprestation_observables}

 This section discusses the physical implications of the numerical results.  The wavefunction comparison shows that, for $\alpha\gg 1$, attaching three vortices to all $N$ particles is energetically less favorable than placing $2N$ vortices in the system center and one on each particle.  Such large vortex states have been the subject of intense interest in the literature (For reviews, see Refs.~\onlinecite{reimann:2002,saarikosk:2010,bloch:2008}).  Here the formation of a vortex is non-trivial because it minimizes interaction energy only.  There is no kinetic energy in Eq.~(\ref{spinorbithfb}).  This energetic competition is akin to the non-trivial competition between inhomogeneous Wigner crystals and uniform quantum liquids in the fractional QH regime.  
 
 The energetics of vortex attachment are determined by the precise form of the interaction and therefore the basis state functions.  For $\alpha \gg1$ the Bessel functions in $\phi_{m}$ impose an oscillating tail in $r$ that manifests in the interaction.  The tail builds up an interaction energy cost as it runs through the rest of the system.  The central vortex state gains in energy by removing the $m=0$ state because the tail of this state overlaps with the most particles.   The numerical results did not find evidence for a Laughlin-type state for $\alpha \gg1$.  Such uniform density states should be energetically more favorable for monotonically decreasing interactions established by Gaussian-like bases, e.g., $\phi^{\text{D}}$.

\begin{table}[t]
\caption{Columns list, from left to right: the particle number, total angular momentum, interaction energy per particle for the exact state and the central vortex state.  The last column lists the overlap of the exact and the central vortex state.  The exact state was obtained from diagonalizing Eq.~(\ref{spinorbithfb}) and the trial state from Eqs.~(\ref{secondquantizedlaughlin}) and (\ref{corejastrow}) with $p=1$  and $\alpha=100$.  The energies units are $2g_{2D}^{\uparrow,\downarrow}/l_{\text{T}}^{2}$.} 
\centering
\vspace{0.1cm}
\begin{tabular}{|l|c|c|c|c|}
  \hline
  $N$            & M          & Exact Energy          & Trial State Energy & Overlap \\[-.1ex] \hline
  $4$     & 	18 & 0.06919 & 0.07028 & 0.989 \\[-.1ex]\hline
  $4$     & 	19 & 0.06449 & 0.06511 & 0.994 \\[-.1ex]\hline
   $5$    & 	30 & 0.08424 & 0.08531 & 0.985 \\[-.1ex]\hline
  $5$     & 	31 & 0.08062 & 0.08140 & 0.990 \\[-.1ex]\hline
  $6$     & 	45 & 0.09839 & 0.09949 & 0.981 \\[-.1ex]\hline
  $6$     &	 46 &0.09547  &0.09640  & 0.986\\
  \hline
\end{tabular}
\label{tab1}
\end{table}

Two-thirds of the wavefunction vortices accumulated at the system center, rather than on individual particles (as in Laughlin states).  The accumulation of many ($\sim N$) wavefunction vortices in one location suppresses the density.  These macroscopic vortices should be visible in time-of-flight measurements of atomic gas systems \cite{bloch:2008,saarikosk:2010}.  Such observations would imply the ability to control and detect interaction-generated vortices as they relocate in many-body states to minimize interaction energy.

\section{Summary}  
\label{sec_summary}

I introduced an implementation of wavefunction vortex attachment in a general matrix-product representation.  I demonstrated the utility of this formalism by validating wavefunctions constructed to describe a model of trapped 2D atomic Fermi gases in the presence of synthetic Rashba SOC and slow rotation.  The flat band limit led to a large central vortex.

The exact method introduced here allows straightforward validation of Jastrow-based ansatz states in small system sizes.  Small system sizes are valuable in studying states with exponentially decaying correlations, e.g., topological quantum liquids.  To reach larger system sizes, approximations introducing singular value decomposition with tensor network-based algorithms \cite{cirac:2009}, e.g., the density matrix renormalization group \cite{white:1992,schollwoeck:2005}, can be used for scale-up.

I acknowledge support from the ARO (W911NF-12-1-0335) and DARPA-YFA (N66001-11-1-4122).

\appendix
\section*{Appendix: Proof of Equation~(\ref{secondquantizedJastrow})}
\label{appendix_jastrow}

This section proves that the second-quantized form of the Jastrow factor, Eq.~(\ref{secondquantizedJastrow}), derives from the first-quantized form.  The proof follows a derivation of Shiota's formula \cite{difrancesco:1994} (written in first quantization) and then second quantizes the operators in this formula.  

I start with the first-quantized Jastrow factor:
\begin{eqnarray}
\mathcal{J}^{2p}_{\gamma=1}=\prod_{j<k}^{N}\left ( T^{\dagger}_{j} -  T^{\dagger}_{k}\right )^{2p}, 
\nonumber
\end{eqnarray}
where, without loss of generality, I choose $\gamma=1$.  The translation operators are understood to act on a suitably chosen state.  I have also assumed that the translation operators act as ladder operators on a basis $\phi_{n}$ that has been mapped to a 1D graph so that $T^{\dagger} \phi_{n} \propto \phi_{n+1}$.   

By defining a new operator:
\begin{eqnarray}
\Pi \equiv\prod_{j,k=1}^{N}\left[1-\epsilon \left ( T^{\dagger}_{j} -  T^{\dagger}_{k}\right )^{p} \right],
\nonumber
\end{eqnarray}
where $\epsilon$ is a small number, I can rewrite the Jastrow factor as:
\begin{eqnarray}
\mathcal{J}^{2p}_{\gamma=1}=(-1)^{pN(N-1)/2}\text{Coeff}_{\epsilon^{N(N-1)}}\left [ \Pi \right],
\nonumber
\end{eqnarray}
where $\text{Coeff}_{\epsilon^{N(N-1)}}$ indicates the coefficient of the $\epsilon^{N(N-1)}$ term in the expansion.  $\Pi$ can then be written in terms of sums over translation operators using the binomial theorem:
\begin{eqnarray}
\Pi &=& \exp\left[ \sum_{i,j=1}^{N} \log \left \{ 1-\epsilon \left ( T^{\dagger}_{i} -  T^{\dagger}_{j}\right )^{p} \right \} \right ] \nonumber \\
&=& \exp\left[- \sum_{ n=1}^{\infty} \frac{\epsilon^{n}}{n}\sum_{l=0}^{np} {np \choose l}  \sum_{i,j=1}^{N} 
\left( T^{\dagger}_{i}\right)^{np-l} \left( -T^{\dagger}_{j}\right)^{l} \right ] \nonumber\\
&=& \exp\left[- \sum_{ n=1}^{\infty} \frac{\epsilon^{n}}{n}\sum_{l=0}^{np} {np \choose l}  (-1)^{l} t_{np-l}t_{l}\right ],
\nonumber
\end{eqnarray}
where the sum over all translation operators is:
\begin{eqnarray}
t_{l}\equiv\sum_{j=1}^{N} \left( T^{\dagger}_{j}\right)^{l}.
\nonumber
\end{eqnarray}
This form for $\Pi$ shows that $\mathcal{J}^{2p}_{\gamma=1}$ can be rewritten in terms of a simple product over operator sums $t_{l}$.

To find the $\epsilon^{N(N-1)}$ coefficient, the exponential can be expanded:
\begin{eqnarray}
\Pi &=& \sum_{k=0}^{\infty}\frac{1}{k!}\left [- \sum_{ n=1}^{\infty} \frac{\epsilon^{n}}{n}\sum_{l=0}^{np} {np \choose l}  (-1)^{l} t_{np-l}t_{l} \right ]^{k}.
\nonumber
\end{eqnarray}
By substituting $\Pi$ into the equation for $\mathcal{J}_{\gamma=1}^{2p}$, the sums over $k$ and $n$ become finite because there are only a finite number of terms contributing to the pre-factor of $\epsilon^{N(N-1)}$:
\begin{eqnarray}
\mathcal{J}_{\gamma=1}^{2p}&=&(-1)^{\frac{pN(N-1)}{2} }\sum_{k=1}^{N(N-1)}\frac{(-1)^{k}}{k!}\times \nonumber \\
& &\hspace{-1.5cm}  \sum_{\substack{n_{1}\geq1,...,n_{k}\geq1 \\n_{1}+...n_{k}=N(N-1)}} \prod_{i=1}^{k} \frac{1}{2n_{i}}\sum_{l=0}^{n_{i}p} {n_{i}p \choose l}  (-1)^{l} t_{n_{i}p-l}t_{l}.
\nonumber
\end{eqnarray}
This form for $\mathcal{J}^{2p}_{\gamma}$ is written with operator sums and can therefore be rewritten in terms of operators in Fock space.  

The aboveJastrow factor can now be rewritten in second quantization.  I rewrite the operators $t_{l}$ in terms of the field operators $\hat{\Psi}({\bf r})$:
\begin{eqnarray}
\hat{\mathcal{J}}_{\gamma}^{2p}&=&(-1)^{\frac{pN(N-1)}{2} }\sum_{k=1}^{N(N-1)}\frac{(-1)^{k}}{k!}\times \nonumber \\
& &\hspace{-1.5cm}  \sum_{\substack{n_{1}\geq1,...,n_{k}\geq1 \\n_{1}+...n_{k}=N(N-1)}} \prod_{i=1}^{k} \frac{1}{2n_{i}}\sum_{l=0}^{n_{i}p} {n_{i}p \choose l}  (-1)^{l} \hat{M}_{n_{i}p-l}\hat{M}_{l},\nonumber
\end{eqnarray}
where the $t_{l}$ operators have become:
\begin{eqnarray}
\hat{M}_{l}\equiv\int d{\bf r} \hat{\Psi}^{\dagger}({\bf r}) \left( T^{\dagger}\right)^{l} \hat{\Psi}({\bf r}).
\nonumber
\end{eqnarray}
Here $ \left( T^{\dagger}\right)^{l} \hat{\Psi}({\bf r}) $ implies translation of orthonormal single-particle basis states in a decomposition of the field operators.  This can be seen by expanding the field operators in terms of the basis states $\phi_{n}$ explicitly:
\begin{eqnarray}
\hat{M}_{l}&=&\sum_{n,n'}\langle \phi_{n}\vert  \left( T^{\dagger}\right)^{l} \vert \phi_{n'} \rangle \hat{c}^{\dagger}_{n}\hat{c}^{\vphantom{\dagger}}_{n'}   \nonumber \\
 &=&\sum_{n,n'}\gamma_{n,n'}^{l} \hat{c}^{\dagger}_{n}\hat{c}^{\vphantom{\dagger}}_{n'}, 
 \nonumber
\end{eqnarray}
where $\hat{c}^{\dagger}_{n}$ creates a fermion in the state $\phi_{n}$. Note that by changing the form for $\gamma_{n,n'}^{l}$, the $\gamma=0$ limit can be obtained.  The matrices defined by $\hat{M}_{l}$ are essentially single-particle density matrices.  The above derivation was specified to a Jastrow factor written for $N$ single component particles with basis states that can be mapped to a 1D graph.  The above derivation can be generalized to multicomponent bases and other graphs, $\Lambda$.

\end{document}